# Robustness of supply chain networks against underload cascading failures

Qihui Yang[a,*], Caterina M. Scoglio[a], Don M. Gruenbacher[a]

[a] Department of Electrical and Computer Engineering, Kansas State University, Manhattan, Kansas, 66506, USA

*Corresponding author
E-mail: qihui@ksu.edu

# Abstract

In today's global economy, supply chain (SC) entities have become increasingly interconnected with demand and supply relationships due to the need for strategic outsourcing. Such interdependence among firms not only increases efficiency but also creates more vulnerabilities in the system. Natural and human-made disasters such as floods and transport accidents may halt operations and lead to economic losses. Due to the interdependence among firms, the adverse effects of any disruption can be amplified and spread throughout the systems. This paper aims at studying the robustness of SC networks against cascading failures. Considering the upper and lower bound load constraints, i.e., inventory and cost, we examine the fraction of failed entities under load decrease and load fluctuation scenarios. The simulation results obtained from synthetic networks and a European supply chain network [1] both confirm that the recovery strategies of surplus inventory and backup suppliers often adopted in actual SCs can enhance the system robustness, compared with the system without the recovery process. In addition, the system is relatively robust against load fluctuations but is more fragile to demand shocks. For the underload-driven model without the recovery process, we found an occurrence of a discontinuous phase transition. Differently from other systems studied under overload cascading failures, this system is more robust for power-law distributions than uniform distributions of the lower bound parameter for the studied scenarios.

# 1. Introduction

Cascading failures and network robustness have been studied extensively in real-world networks such as power systems and traffic networks [2–9]. Failures in these models are often overload-driven. Pahwa et al. [9] built a simple but realistic overload cascade model for power systems and studied its emergent behavior against power outages, which can lead to the load redistribution across the network and result in more failures due to the power flows exceeding the line capacity. They observed a sudden breakdown of the system with an increased load level and a large network size. Many works study the cascading failure phenomenon from a single network perspective, and there have also been recent studies of failure cascades in interdependent networks. Buldyrev et al. developed in [10] a one-to-one correspondence model to study the robustness of interconnected networks against cascading failures, and found that removal of a critical fraction of nodes could lead to a complete breakdown of the system. Since the work [10], many works have studied the robustness of interdependent networks from various perspectives [11–14], in which disruptions of components in one system may propagate and cause elements in the other system to fail.

In the competitive economic market, SC entities often build business relationships with outsourcing partners to reduce the overall cost and promote productivity. As a result, SCs have become more complicated and geographically dispersed, increasing the frequencies of SC disruptions [15]. Due to the increased dependencies among entities, disruptions of a company's operation can result in revenue losses in its business partners, and cascade to other components in the SC [16]. On the other hand, sustainability concerns have pushed for higher efficiency in the use of resources and reduction in protective redundancies in SCs, thus making SCs more susceptible to disruptions [17]. SC entities may suffer from financial losses due to delays in the flow of goods caused by natural disasters or intentional attacks. For example, heavily relying upon transportation services for timely delivery of animals among ranches, stockers, feedlots, and meat-processing plants, the beef industry can suffer substantial economic losses due to disruptions in transportation infrastructure caused by natural disasters and movement restrictions during disease outbreaks [12]. During the 2011 Japan earthquake and tsunami, Toyota Motor Company suffered at least a 140,000-vehicle production loss. This adverse effect spread to other countries, which led to a massive collapse in the global automotive and electronics industry [18]. In the following weeks after the disaster, Toyota in North America experienced shortages of over 150 parts, resulting in curtailed operations at only 30% of capacity [19].

In light of these low probability and high impact disruptions, several studies have utilized the above overload failure models to analyze the SC robustness against cascading failures [20,21]. From a complex network perspective, a supply chain is also termed a supply network, in which network nodes and links refer to the SC entities and supply-demand relationships between the entities, respectively. Tang et al. [21] examined the robustness of an interdependent SC network model, which consists of an undirected cyber layer and a directed physical layer, subject to different node removal strategies. However, the nature of cascading failures in SC systems is mostly underload-driven. When entities cannot fulfill the expected production requirement to overcome the fixed production costs, they will fail to gain profit and possibly exit the market. When a node is disrupted, its downstream and upstream neighbors will be affected due to supply shortage and demand losses, respectively. If the neighbor node's remaining load drops below the lower bound constraint, i.e., cost, new failure occurs, and cascades to other nodes in the entire system [22,23]. Tang et al. [24] assessed the SC system robustness in the form of production capability losses, but they assumed that the failed node loads only propagate downstream without consideration of mitigation strategies. Wang et al. [22] attempted to improve the cluster SC network resilience against cascading failures, leveraging insights from the ant colony's spatial fidelity zones. In the model developed by Wang et al., a node can propagate the failure impact both upstream and downstream, and can dynamically change the strength of the business relationship with its neighbors [23]. To the best of the authors' knowledge, there are few works considering underload-driven failures, and this work attempts to add a new element to this domain with a focus on the phase transition behavior by incorporating network flows.

Given the tremendous damages caused by the disruptions, understanding the nature of the systemic failure in SC that goes beyond a single component behavior is a significant problem to be addressed. The goal of this work is to build an underload cascade failure model and analyze its behavior against disruptive events. First, we will build a generalized underload cascade failure model with and without recovery strategies. The material flow of goods through an entity node defines its load. Second, we will examine the fraction of failed nodes in the system under scenarios of load decrease and load fluctuations. Third, we will provide analytic results based on the assumption of equal load redistribution upon failures for the studied scenarios.

The contributions of this paper are: (i) Using the proposed model with recovery strategies, we confirm that strategies of backup suppliers and surplus inventories, often adopted in actual SCs, improve the system robustness. In addition, the system is relatively robust against load fluctuations but is more fragile to load decrease. (ii) Without the recovery process, we found numerically and analytically a discontinuous phase transition of the system under a load decrease scenario. Based on statistical physics, this indicates that a small fraction of failures can result in a sudden breakdown in the SC system. More specifically, the model is more robust under the power-law distribution than the uniform distribution of the lower bound load parameter for the studied scenarios. These findings indicate that underload driven systems have different behaviors against cascading failures compared with overload cascade models, and need to be further explored.

## 2. Underload cascading failure model

### 2.1 Failed node load propagation in supply chain networks

In SC networks, featured with multi-hierarchy, nodes are entities and edges denote the business relationships between the entities. SC entities are both demand-side and supply-side, and those in the same tier have similar network connections, core businesses, and competitive environments [25–27]. In this work, the sum of material flows that go through an entity node defines its load $L$ ($t^{-1}$), i.e., the total number of products that an entity has sold per time unit. Accordingly, the sum of incoming flows always equals the sum of outgoing flows for each node $i$, i.e., $\sum_{j \epsilon \Gamma_i^U(t)} F_{ji}(t) = \sum_{j \epsilon \Gamma_i^D(t)} F_{ij}(t)$, with demand and supply balanced. $\Gamma_i^U(t)$ and $\Gamma_i^D(t)$ are the set of upstream nodes and set of downstream nodes connected to node $i$, respectively.

Each node $i$ has a specific upper bound load $A_i$ and lower bound load $B_i$, which are proportional to the initial node load $L_i(0)$, and are given by $A_i = aL_i(0)$ ($a > 1$) and $B_i = bL_i(0)$ ($0 < b < 1$), where $a$ and $b$ are upper and lower bound parameters, respectively. More specifically, the load of node $i$ at time $t$, $L_i(t)$, is always below its $A_i$, which is the total number of products an entity currently has, i.e., inventory. Meanwhile, in normal conditions, an entity's load should be above $B_i$, reflecting costs such as labor cost and maintenance fee. Residual load of a node $i$ at time $t$, given by $RL_i(t) = A_i - L_i(t)$, indicates the additional available products an entity can provide, i.e., surplus inventory. Under disruptions, a node fails if its load $L_i(t)$ falls below $B_i$, meaning that the entity fails to gain profit in the competitive market.

When an SC network suffers from disruptive events, some nodes could get underloaded and fail initially. Caused by the connections between the entities, the impact of these initial failures could result in more failures and spread to the entire system. Fig.1 shows a simplified example of the load propagation process in a four-tier SC network. When nodes 9 and 13 fail at time $t$, loads of the failed nodes first propagate to their neighbor nodes 4, 14, 15 and 8, 18 through connectivity links. Then, these affected nodes 14, 15 further impact their downstream nodes 18, 19 and 20 by reducing supply. Meanwhile, node 8 influences its upstream neighbor nodes 2, 3 and 4 by cutting back demand.

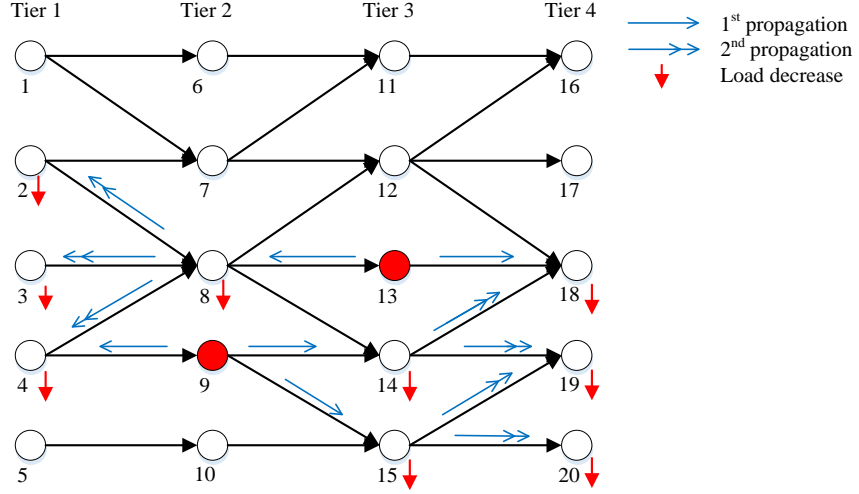

**Fig. 1.** Failed node load propagation in a four-tier supply chain. Red circles represent the initial failures caused by load decrease or fluctuations.

## 2.2 Cascading failure process

In this section, the cascading failure processes with and without recovery measures are described with the following steps. Step 0: We calculate all the initial node loads $L_i(0)$, load constraints for nodes, and initial flows on the links. Keeping the load constraints for all nodes unchanged, we decrease or fluctuate the initial node loads to simulate initial failures. The new initial load of node $i$ after load decrease or fluctuations is denoted as $L'_i(0)$. Step 1: The new failure propagates throughout the system, and all the loads and flows affected get updated. Step 2: With recovery measures, the nodes affected can mitigate losses by altering flows with existing partners or building new partners. Then, if there are still nodes with load below their lower bound load, new failure occurs and we continue the process from Step 1–2. Such a procedure is repeated until no new failure happens. More details are illustrated as follows.

### 2.2.1 Model initialization

In this step, we first calculate the node degree for all nodes and weight on all edges. Node degree of node $i$ is obtained by summing its indegree $d_i^+$ and outdegree $d_i^-$, i.e. $d_i = d_i^+ + d_i^-$. Reflecting the business relationship strength between nodes $i$ and $j$, weight on edge $e_{ij}$ is calculated by $w_{ij} = (d_i \cdot d_j)^\theta$, with $\theta = 0.5$ [22,23].

Then, we compute the initial flows, initial node loads and load constraints in the network. Suppose the SC network has $T$ number of tiers, and the initial loads for nodes in the last tier $T$, equivalent to the demand for final customers, are known and equal. We rescale the weight on incoming edges of node $j$ such that new weights on incoming edges of node $j$ sums to 1, i.e., $\sum_{i \in \Gamma_j^U} w'_{ij} = 1$:

$$w'_{ij} = w_{ij} / \sum_{k \in \Gamma_j^U} w_{kj} \tag{1}$$

where $\Gamma_j^U$ is the set of upstream nodes connected to node $j$ and $\sum_{k \in \Gamma_j^U} w_{kj}$ is the sum of weights on incoming edges of node $j$.

Given initial node loads in tier $T$, we calculate the initial flow over $e_{ij}$ between node $i$ in tier $T-1$ and node $j$ in tier $T$ by

$$F_{ij}(0) = w'_{ij} \cdot L_j(0) \quad (2)$$

Accordingly, the load of node $i$ in tier $T-1$ is the sum of the outgoing flows of node $i$

$$L_i(0) = \sum_{k \in \Gamma_i^D} F_{ik}(0) \quad (3)$$

where $\Gamma_i^D$ is the set of downstream nodes connected to node $i$.

Following the same method, we sequentially calculate the initial loads of nodes in tier $T-2, T-3, \cdots, 2, 1$. After the determination of all nodes' initial loads, we calculate the load constraints $A_i$ and $B_i$ of each node $i$.

### 2.2.2 Failed node load propagation

We assume that when a node fails, it can neither receive supplies from upstream neighbors, nor ship products to its downstream partners. Once a node $i$ fails at time $t$, both its load and the flows on its incoming and outgoing edges are set to 0. The impact of failed node $i$ first propagates upstream using Eq. 4, in which we calculate the loss of each node $j$ in its upstream neighbor set, $\Gamma_i^U$, suffers at time $t$, i.e., $\Delta L_j(t)$, and update the corresponding loads and flows.

$$\begin{cases} \Delta L_j(t) = L_i(t-1) F_{ji}(t-1) / \sum_{m \in \Gamma_i^U} F_{mi}(t-1) \\ L_j(t) = L_j(t-1) - \Delta L_j(t) \\ F_{ji}(t) = F_{ji}(t-1) - \Delta L_j(t) \end{cases} \quad (4)$$

Then, affected nodes $j$ further impact nodes $k$ in its upstream neighbor set, $\Gamma_j^U$ using Eq. 5.

$$\begin{cases} \Delta L_k(t) = \Delta L_j(t) F_{kj}(t-1) / \sum_{m \in \Gamma_j^U} F_{mj}(t-1) \\ L_k(t) = L_k(t-1) - \Delta L_k(t) \\ F_{kj}(t) = F_{kj}(t-1) - \Delta L_k(t) \end{cases} \quad (5)$$

Similarly, we simulate the failure propagation of node $i$ by updating the loads and flows for each node $j$ in $\Gamma_i^D$ and then for each node in $\Gamma_j^D$. This process continues to tier 1 and $T$, mimicking the ripple effect spreading throughout the system.

### 2.2.3 Load recovery process

In this step, the nodes which have not been marked as failed can recover their loads by requesting rush orders or building new business partnerships to mitigate losses. Since acquiring new partners will bring additional costs, we assume node $i$ will first select from surviving nodes in its upstream neighbor set $\Gamma_i^U$ and downstream neighbor set $\Gamma_i^D$. If all the neighbor nodes still cannot help it recover above $B_i$, node $i$ will build links with surviving nodes in its upstream non-neighbor set $\bar{\Gamma}_i^U(t)$ and downstream non-neighbor set $\bar{\Gamma}_i^D(t)$.

According to the node load definition, each node's demand and supply should be balanced. When node $i$ alters its outgoing flow with node $j$ in $\Gamma_i^D$, i.e., $F_{ij}(t)$, loads of both nodes $i$ and $j$ need to be updated with the same increase, $\psi_i(t) = \psi_j(t)$. Meanwhile, node $i$ needs to reconfigure its incoming flows with nodes in $\Gamma_i^U$ such that the sum of outgoing flows equals the sum of incoming flows through node $i$. As a result, the sum of load increase in each tier $s$, $\sum_{i \in tier\ s} \psi_i(t)$, should be equal.

On the other hand, as the numbers of surviving nodes are not necessarily the same for each tier, the load loss of all the surviving nodes in each tier, $\sum_{i \in tier\ s}(L'_i(0) - L_i(t))$, can be different. In addition, each node should at most provides its residual, i.e., $RL_i(t) \geq 0$, but the sum of residual load for nodes in each tier $s$, $\sum_{i \in tier\ s} RL_i(t)$, can be different.

Due to the reasons above, we identify the tier $\gamma_a$ with the minimum residual load $RL_{MIN}$ and tier $\gamma_b$ with the maximum loss $\Delta L_{MAX}$ (Eq. 6). In each tier, the load increase $\psi_i(t)$ allocated to each node $i$ should sum up to the smaller value of $RL_{MIN}$ and $\Delta L_{MAX}$, which is defined as the load increase objective for each tier, i.e., $Q = \sum_{i \in tier\ s} \psi_i(t)$. Once each tier reaches the load increase objective $Q$, the recovery process stops.

$$\begin{cases} RL_{MIN} = \min_s[\sum_{i \in tier\ s} RL_i(t)], s = 1,2,\ldots,T \text{ and } \gamma_a = s^* \\ \Delta L_{MAX} = \max_s[\sum_{i \in tier\ s}(L'_i(0) - L_i(t))], s = 1,2,\ldots,T \text{ and } \gamma_b = s^* \\ \sum_{i \in tier\ s} \psi_i(t) = \min(RL_{MIN}, \Delta L_{MAX}), s = 1,2,\ldots,T \end{cases} \quad (6)$$

More specifically, if $RL_{MIN} < \Delta L_{MAX}$, we first allocate load increase $Q$ to the nodes in tier $\gamma_a$. Each node $i$ in tier $\gamma_a$ increases its load by $\psi_i(t) = RL_i(t)$ and distributes this load increase to its suppliers in $\Gamma_i^U$. Otherwise, the load increase $Q$ is first allocated to the nodes in tier $\gamma_b$, and each node in tier $\gamma_b$ distributes the load increase $\psi_i(t) = L'_i(0) - L_i(t)$ to its suppliers in $\Gamma_i^U$. If the first selected supplier $j$'s $RL_j(t)$ cannot satisfy the load increase request of node $i$, node $i$ will continue to ask from the next supplier in $\Gamma_i^U$. If $\sum_{j \in \Gamma_i^U} \psi_j(t) < \psi_i(t)$, node $i$ will build links with nodes in non-neighbor set $\bar{\Gamma}_i^U(t)$, until objective $\psi_i(t)$ is met. Like the inverse of the load propagation process, this process continues until all related suppliers in tier 1 have increased their loads. Similarly, loads of surviving nodes downstream get updated.

## 3. Numerical simulation results

Researchers often study the phase transition behavior exhibited by the system through stressing external forces to it until the rupture point [28]. In this section, we examine the robustness of the SC networks under the stress of load decrease and fluctuations. Note that each node's load constraints, $A_i$ and $B_i$, remained fixed as strength of the stress $\delta$ or $\sigma$ changes under each scenario.

As the failures are underload-driven, we conduct extensive simulations with lower bound parameter $b$ following (i) uniform and (ii) power-law distribution, which are two commonly used families of distributions [9,29]. Uniformly distributed over $[b_{min}, b_{max}]$, denoted by $U[b_{min}, b_{max}]$, the probability density function for a random variable $b$ is given by $p(b) = 1/(b_{max} - b_{min}) \cdot 1_{b_{min} \leq b \leq b_{max}}$. Following a power-law distribution, the probability density function for a random variable $b$ is of the form $p(b) = k \cdot (b)^{-\gamma}$ with $b \in [b_{min}, 1]$.

### 3.1 Synthetic networks

Hernández and Pedroza-Gutiérrez [30] constructed random network models in bipartite graphs to model the theoretical SC networks. Similarly, we generate synthetic networks for a four-tier SC network, as shown in Fig. 1, representing suppliers, production centers, distribution centers, and customers. $N$ nodes are generated and equally divided into four tiers, in which one enterprise node belongs only to one tier. Links are created randomly with a given connection probability $p$, which is the likelihood of existing business relationships between nodes in two tiers. Additional efforts are made to ensure the network created is without self-loops. Despite the random placement of links, most nodes will have approximately the same number of connections. For example, the average node outdegree of a network with 100 nodes in each tier ($N$=400) and $p$=0.1 will be around 10.

### 3.1.1 Load decrease

In this scenario, we consider a negative demand shock, in which demand for goods or services shrinkages suddenly. To model the load decrease, we simultaneously decrease the initial loads for all nodes in tier 4 by a factor $\delta$, i.e., $L'_i(0) = (1-\delta)L_i(0)$, and then calculate all the flows and node loads in tiers 1–3. Keeping the network topology unchanged, it is equivalent to a uniform decrease of all the nodes by a factor $\delta$. In each realization, $\delta$ changes from 0 to 1 with a step size of 0.02. We record the fraction of failed nodes $f$ at the end of the simulation, and the results are averaged over 100 realizations with network size $N$=400 and connection probability $p$=0.1.

In Fig. 2(a)–(c), we found a sharp collapse of the system when there are no recovery measures and $b$ is uniformly distributed over $[b_{min}, b_{max}]$. More specifically, the critical point above which the failure occurs is only determined by $b_{max}$, the upper limit of the uniform distribution. For example, in Fig. 2(b), in which $B_i$ ranges between $[0.2L_i(0), 0.7L_i(0)]$, when $\delta > 0.3$, i.e., $L'_i(0) < 0.7L_i(0)$, initial failures start to appear and will propagate throughout the system, resulting in the collapse of the whole system.

When recovery measures are included and $b \in U[b_{min}, b_{max}]$, results in Fig. 2(a)–(c) show that the recovery strategy can reduce the scale of systemic failure, as the reconfiguration of the trade flows among alive nodes can absorb losses of nodes affected by the initial failures. In Fig. 2(b), when $\delta > 0.8$, i.e., $L'_i(0) < 0.2L_i(0)$, all the loads fail at the beginning and are marked as failed, so they will not mitigate losses under the recovery process.

With $b$ in the form of power distribution, the system collapses when $\delta$ increases to 0.88 without recovery process in Fig. 2(d), indicating that the system becomes more robust compared to the uniform distribution case.

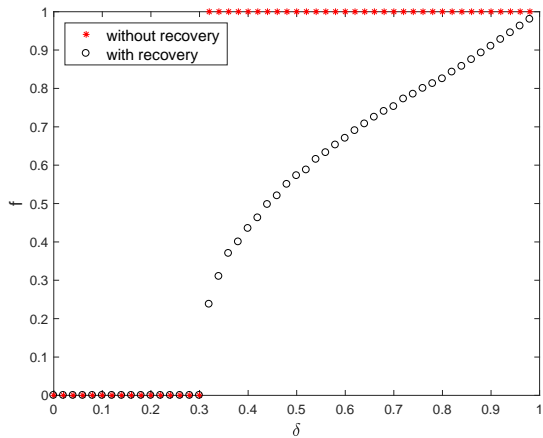
(a) $b \in U[0, 0.7]$

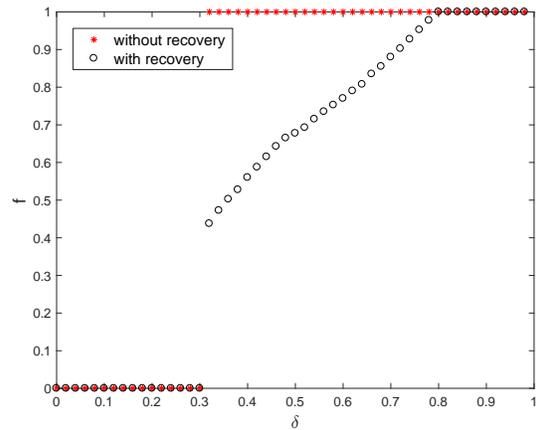
(b) $b \in U[0.2, 0.7]$

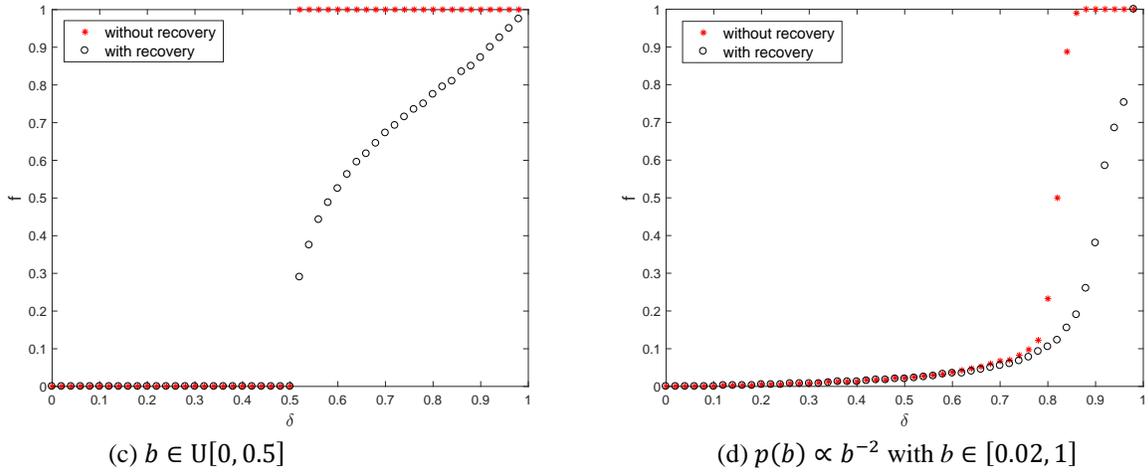

(c) $b \in U[0, 0.5]$      (d) $p(b) \propto b^{-2}$ with $b \in [0.02, 1]$

**Fig. 2.** Plots of the fraction of failed nodes $f$ versus the relative load decrease $\delta$ mimicking the demand shock.

### 3.1.2 Load fluctuations

We mimic the load fluctuations by setting final customers' new initial load as $L'_i(0) = (1 + \sigma \xi_i)L_i(0)$, i.e., $L'_i(0) \in [(1 - \sigma)L_i(0), \ (1 + \sigma)L_i(0)]$, where $\xi_i$ is a random variable uniformly distributed in [-1, 1]. Then, we calculate the new initial loads for the nodes in tier 1–3 and flows on the edges. It is equivalent to allowing all the initial loads to fluctuate by a fraction of $\sigma$. When $\sigma$ varies between [0, 1], upper bound parameter $a$ is set to be two to ensure $L'_i(0) < A_i$, with $A_i = aL_i(0)$. Results are averaged over 100 runs of the simulation, with network size $N$=400 and connection probability $p$=0.1.

Fig. 3 shows the changes in the fraction of failed nodes as variation size $\sigma$ increases. Compared with the load decrease scenario, the system collapse happens less abruptly. When recovery measures are not included and $b$ is uniformly distributed over [0, 0.9], the fraction of failed nodes reaches a plateau of around 80% as fluctuation escalates. In comparison, when $b$ is uniformly distributed over [0, 0.8], about 25% of nodes failed in the end, suggesting that the system is relatively robust.

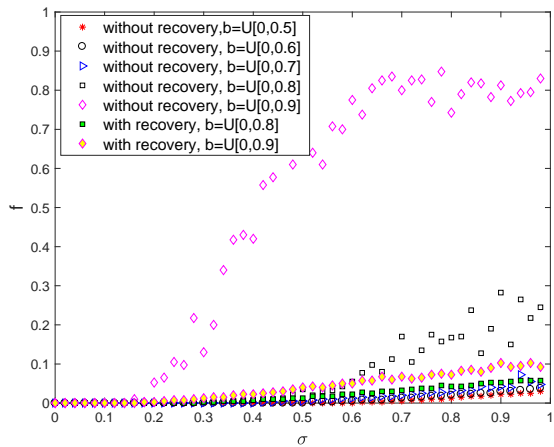

**Fig. 3.** Plot of the fraction of failed nodes $f$ versus the relative strength of load fluctuations $\sigma$.

## 3.2 Real-world example

The underload cascading model proposed in this work is independent of the network topologies and can be applied to other types of SC networks. In this section, we apply the proposed model to a European supply chain network obtained from [1], which includes 11 raw materials suppliers, 3 plants, 5 warehouses, and 18 markets (Fig. 4).

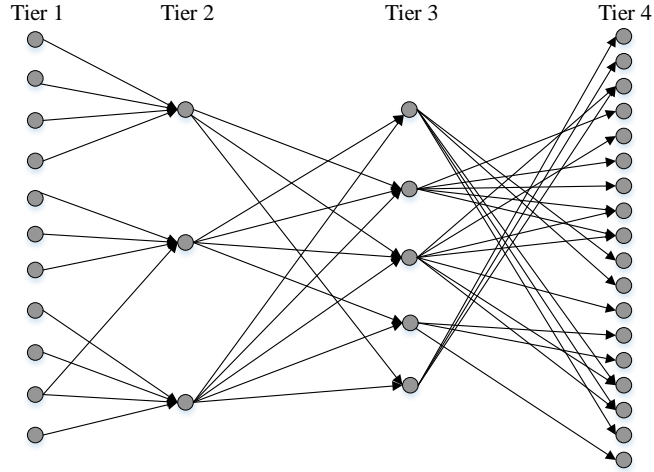

**Fig. 4.** Network structure of a European supply chain (from Cardoso et al. [1]).

As shown in Fig. 5, simulation results under load decrease/fluctuation scenarios coincide with previous simulation results obtained from synthetic networks. Regarding load decrease, we observe a discontinuous phase transition for the system without recovery measures, and this system becomes more robust with the addition of recovery measures. Also, the system is relatively robust in the case of load fluctuations. For instance, without recovery measures, the fraction of failed nodes varies between 40%–50% when the relative strength of load fluctuations $\sigma$ is larger than 0.66.

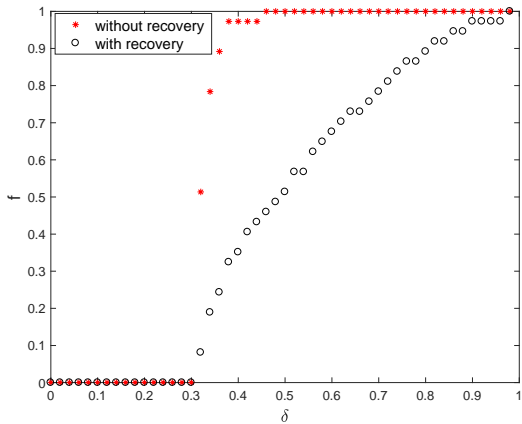

(a) Load decrease scenario with $b \in U[0, 0.7]$

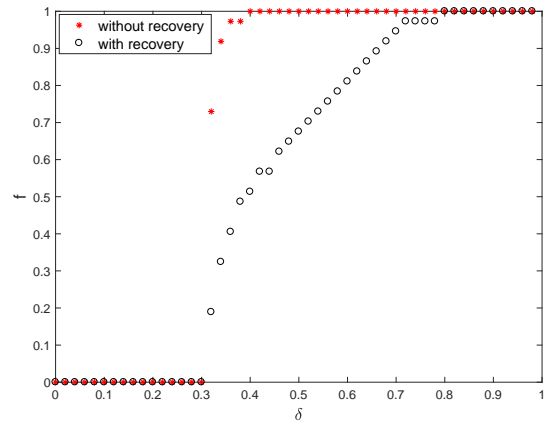

(b) Load decrease scenario with $b \in U[0.2, 0.7]$

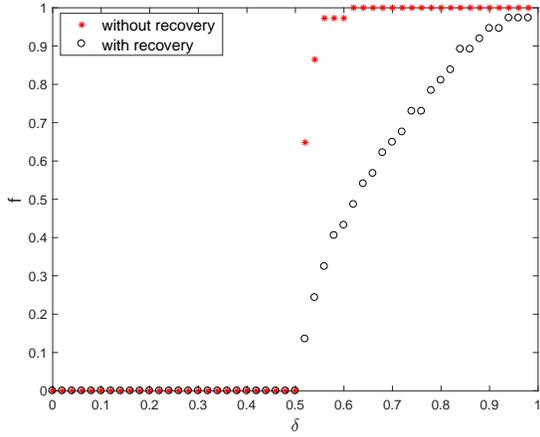
(c) Load decrease scenario with $b \in U[0, 0.5]$

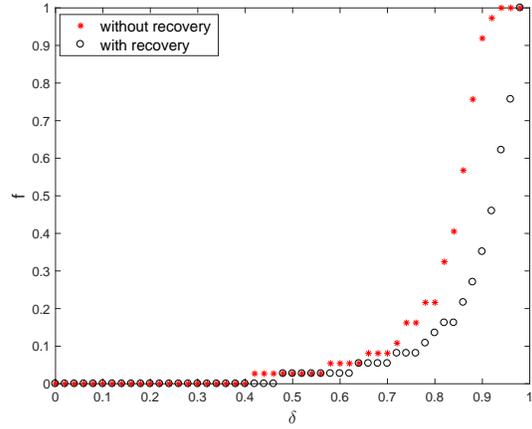
(d) Load decrease scenario with $p(b) \propto b^{-2}$ and $b \in [0.02, 1]$

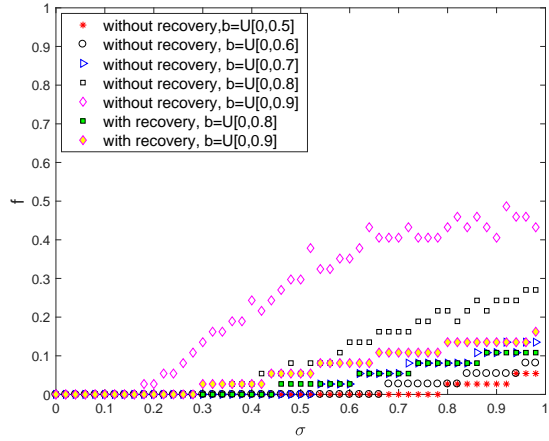
(e) Load fluctuation scenario
**Fig. 5.** Simulation results using the European supply chain network.

## 4 Mean-field analysis

In this section, we forecast a discontinuous transition for the cascading failure model without the recovery process using mean-field analysis. In power systems, power flows can redistribute in the whole system upon failures according to the Kirchhoff's law, which is dependent on power line impedance. Similarly, the effects of disruptions can spread to the entire SC network, and the flow redistribution in the system is dependent on the business relationships among entities. This feature inspires us to leverage the equal load redistribution model that has been used in power systems [9,29,31,32]. The assumption is that when a node fails, the load it carries before the failure will be redistributed equally among all the remaining nodes. The equal load redistribution assumption is originated from the widely used democratic fiber bundle model [33], in which *N* parallel fibers with different failure capacity share an applied force equally.

In the following, we analyze the load decrease scenario using a simple equal-load redistribution model. Suppose there are *N* nodes with a lower bound load $B_i$ characterized by a probability distribution $p(B)$, and failures happen in discrete time steps *t*=0, 1… The fraction of failed nodes and the number of surviving nodes until cascade stage $t$ is denoted as $f_t$ and $N_t$ respectively. When the load of a node goes below $B_i$, the node fails and its load gets redistributed equally among the remaining surviving nodes.

Suppose all the nodes initially carried the same load $\bar{L}'_0$. There is no failure before the disruptions, thus $f_0 = 0$ and $N_0 = N$. Under disruptions, a fraction of nodes $f_1 = \int_{\bar{L}'_0}^{\infty} p(B)dB$ immediately fails, since their load $\bar{L}'_0$ is below the lower bound load. After the first stage, the number of surviving nodes equals to $N_1 = (1 - f_1)N$, and the new load per node becomes $\bar{L}_1 = \bar{L}'_0 - \frac{f_1 \bar{L}'_0 N}{(1-f_1)N} = \left(1 - \frac{f_1}{1-f_1}\right)\bar{L}'_0$. The cascade failure process continues recursively, and the mean-field equations for the $(t+1)^{th}$ stage are as follows:

$$\begin{cases} f_{t+1} = \int_{\bar{L}_t}^{\infty} p(B)dB \\ N_{t+1} = (1 - f_{t+1})N \\ \bar{L}_{t+1} = \bar{L}_t - (f_{t+1}N - f_t N)\bar{L}_t/N_{t+1} = [1 - (f_{t+1} - f_t)/(1 - f_{t+1})]\bar{L}_t \end{cases} \quad (7)$$

where $(f_{t+1}N - f_t N)/N_{t+1}$ is $\frac{\text{Number of lines that survive stage } t \text{ but fail at } t+1}{\text{Number of lines that survive stage } t+1}$.

Eq. 7 can be simplified as

$$f_{t+1} = F(\bar{L}'_0 \prod_{t=1}^{t}(1 - \frac{f_t - f_{t-1}}{1-f_t})) \quad (8)$$

where $F(x) = \int_x^{\infty} p(B)dB$.

From Eq. 8, we can see that the critical point $f^*$ is mainly dependent on the distribution of $B_i$. To obtain the fraction of failed nodes, we provide analytic solutions by numerically solving Eq. 8 and verify them by simulating the above equal-load redistribution process under disruptions. We assume the initial node load $\bar{L}_0 = 1$ before the disruptions, and thus lower bound load $B_i = b\bar{L}_0 = b \cdot 1$ with $b \in [b_{min}, b_{max}]$. Under load decrease, the new initial load becomes $\bar{L}'_0 = (1 - \delta)\bar{L}_0 = (1 - \delta) \cdot 1$.

The results obtained from the simulation are averaged over 100 runs and compared with the results calculated from equations in Fig. 6, in which a discontinuous phase transition occurs in all cases. In the uniform distribution case, the critical point below which a sudden collapse happens is only determined by $b_{max}$. When the new initial load $\bar{L}'_0$ is above $B_i$, there are no failures in the system. Once $\bar{L}'_0$ falls below $B_i$, the system collapses because the fraction of failed node obtained from the equation will increase to 1. When $b$ follows a power-law distribution, the system is more robust, and there is an abrupt breakdown of the system around $\delta \approx 0.9$. Interestingly, this is contrary of the mean-field result for the overload cascade model [9], in which the discontinuous jump happens with no precursors in power-law distributions, and with precursors in line capacity following a uniform distribution. This indicates that the scale of cascade failures in SCs could be significantly affected by the shape of the $b$ distribution, which is closely related to a company's cost.

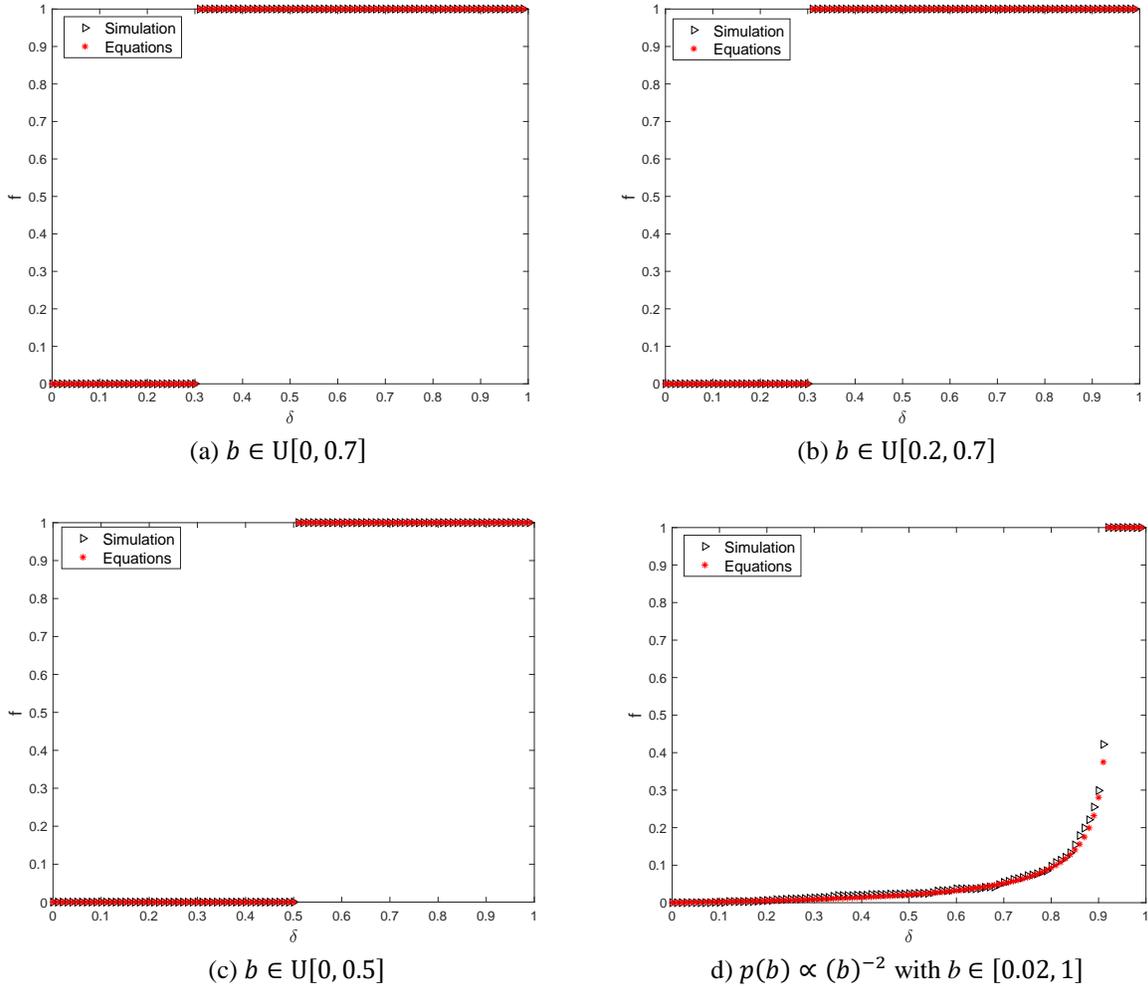

**Fig. 6.** Plots of the fraction of failed nodes *f* as a function of the relative load decrease $\delta$ using the mean-field model.

## 5. Conclusions

In this paper, we constructed an underload cascading failure model to study the robustness of SC networks under load decrease and load fluctuation scenarios. Most real-world SCs are equipped with redundancies such as surplus inventory and backup suppliers, and our simulation results from both synthetic networks and real network topologies show that the recovery strategies can significantly reduce the scale of the systemic failure when disruptive events happen. In addition, the system is relatively robust under the load fluctuation scenario, as the fraction of failed nodes escalates only when variation size $\sigma$ is very high without considering recovery measures, which rarely happens in reality. Compared to the load fluctuations, the system appears more vulnerable against disruptive events such as load decrease, i.e., demand shock, and SC decision-makers need to take proactive protection measures to reduce or avoid the impact of such disruptive events.

Since many cascading failure models do not consider the recovery process, we also studied the behavior of the proposed underload-driven model without the recovery measures. We found that the model exhibits a discontinuous phase transition behavior under load decrease scenario as predicted by the mean-field analysis. More specifically, under different distributions of lower bound parameter *b*, i.e., cost per output, the system is more robust when *b* follows a power distribution compared to the uniform distribution for the studied scenarios. These emergent behaviors observed are different from the analytic results derived from the overload-driven system by Pahwa et al. [9], in which the power-law distribution of capacity results in a more abrupt system breakdown.

In this paper, we qualitatively show the dynamic behavior of specific SC networks against disruptions and do not take the whole complexity of SCs into account. For example, the proposed model disregards the possible internal connections between entities in the same tier. Future work can add more features to the current model and analyze the system behavior under disruptive events. Since the links in the SC synthetic network are randomly generated based on a connection probability and we disrupt all nodes with a factor $\delta$ or $\sigma$, the initial failures triggered in this work can be seen as caused by random attacks. More simulations can be conducted to examine the phase transition behavior of SC systems under targeted attack with various network topologies. In this work, we mainly focus on the robustness assessment, thus designing an optimal recovery strategy is not our primary focus. The recovery process developed assumes that SC entities have full knowledge of the surplus inventory information in the whole system, and future work can include more realistic assumptions regarding mitigation strategies.

# Acknowledgements

This work was supported by the National Science Foundation under Grant Award CMMI-1744812. The contents of the paper do not necessarily reflect the position or the policy of funding parties.